\documentclass[prb,superscriptaddress,floatfix,twocolumn,showpacs,amsmath,amssymb,final]{revtex4}
\usepackage{graphicx} 
\usepackage{dcolumn} 
\usepackage{bm} 

\begin{document}

\title{Local spin spirals in the N\'eel phase of La$_{2-x}$Sr$_x$CuO$_4$}

\author{Andreas L\"uscher}
\affiliation{School of Physics, University of New South Wales, Sydney 2052, Australia}
\author{Gr\'egoire Misguich}
\affiliation{Service de Physique Th\'eorique, CEA Saclay, 91191 Gif-sur-Yvette Cedex, France}
\author{Alexander I. Milstein}
\affiliation{Budker Institute of Nuclear Physics, 630090 Novosibirsk, Russia}
\author{Oleg P. Sushkov}
\affiliation{School of Physics, University of New South Wales, Sydney 2052, Australia}

\date{\today}

\begin{abstract}
Experimental observations of lightly doped La$_{2-x}$Sr$_x$CuO$_4$, $x < 0.02$, revealed remarkable magnetic properties such as the incommensurate noncollinear ordering (additional to the N\'eel ordering) and a tremendous doping dependence of the uniform longitudinal susceptibility. We show that the spiral solution of the $t$-$t'$-$t''$-$J$ model obtained by taking into account the Coulomb trapping of holes by Sr ions describes these puzzling data perfectly well. Our solution firstly explains why the incommensurate structure is directed along the orthorhombic $b$-axis, and secondly allows a numerical calculation of the positions and shapes of the incommensurate neutron scattering peaks. Thirdly, we calculate the doping dependence of the spin-wave gap, and lastly, we study the longitudinal magnetic susceptibility and show that its doping dependence is due to the noncollinearity of the spin spiral.
\end{abstract}


\maketitle

\section{Introduction}
The phase diagram of La$_{2-x}$Sr$_x$CuO$_4$ (LSCO) shows that the magnetic state changes drastically with Sr doping. The three-dimensional antiferromagnetic (AF) N\'eel order identified~\cite{keimer92a} below $325 K$ in the parent compound La$_2$CuO$_4$ (LCO) disappears at doping $x\approx0.02$ and gives way to the so-called spin-glass phase which extends up to $x\approx 0.055$. In both, the N\'eel and the spin-glass phase, the system essentially behaves like an Anderson insulator and exhibits only hopping conductivity. Superconductivity then sets in for doping $x>0.055$ (see Ref.~\onlinecite{kastner98}). One of the most intriguing properties of LSCO is the static incommensurate magnetic ordering at low temperature observed in elastic neutron scattering experiments. Such ordering is a generic feature of LSCO, because it has been observed in the superconducting phase~\cite{yamada98}, in the spin-glass phase~\cite{wakimoto99,matsuda00,fujita02}, and in the N\'eel phase~\cite{matsuda02}. The incommensurate ordering manifests itself as an additional elastic scattering peak shifted with respect to the AF position: ${\bf Q}={\bf Q}_{AF}+\delta{\bf Q}$, where ${\bf Q}_{AF}=(\pi,\pm\pi)$, setting the lattice spacing $a$ equal to one. According to experiments in the underdoped region $0.055 < x < 0.12$, the shift scales linearly with doping and is directed along the crystal axes of the square lattice, $\delta{\bf Q}\approx 2x(\pm\pi,0)$ or $\delta{\bf Q}\approx 2x(0,\pm\pi)$, see Ref.~\onlinecite{yamada98}. In the spin glass phase, $0.02 < x < 0.055$, the shift also scales linearly with doping, but it is directed along the orthorhombic $b$-axis, $\delta{\bf Q}\approx \pm\sqrt{2}x(\pi,-\pi)$, see Refs.~\onlinecite{wakimoto99,matsuda00,fujita02}. Finally, in  the N\'eel phase for $x < 0.02$, the incommensurability is almost doping independent and directed along the orthorhombic $b$-axis (analogous to the spin glass phase), $\delta{\bf Q}\approx \pm 0.02 \sqrt{2}(\pi,-\pi)$, see Ref.~\onlinecite{matsuda02}. Experimental data for the elastic neutron scattering probability at $x=0.01$ and $x=0.014$ are shown in Fig.~3 of Ref.~\onlinecite{matsuda02} (and reproduced in Fig.~\ref{fig:neutronscattering} in this paper). The extracted correlation lengths summarized in Tab.~1 of Ref.~\onlinecite{matsuda02} clearly indicate the presence of long-range incommensurate correlations. The correlation length is about 200 \AA\ along the orthorhombic $b$-direction and more than 500 \AA\ along the $a$-direction. To resolve the small incommensurate peaks in the background of the huge commensurate peak, the authors of Ref.~\onlinecite{matsuda02} used the fact that the scattering amplitude for neutrons interacting with electron spins is of the form
\begin{equation} \label{eq:hint}
T_{\bf q} = {\bm \mu}^{(N)} \cdot {\bf B}_{\bf q} = 4 \pi {\bm \mu}^{(N)} \cdot \left[ {\bm \mu}_{\bf q} - \frac{{\bf q} \left( {\bm \mu}_{\bf q} \cdot {\bf q}\right)}{q^2} \right] \ .
\end{equation}
Here ${\bm \mu}^{(N)}$ is the magnetic moment of the neutron, ${\bf q}$ the momentum transfer and ${\bf B}_{\bf q}$ the Fourier transformation of the magnetic field generated by the magnetization density ${\bm \mu}\left({\bf r}\right)$, which in momentum space becomes ${\bm \mu}_{\bf q}$. It is well known~\cite{kastner98}, that due to the Dzyaloshinski-Moriya (DM) interaction, the commensurate N\'eel magnetization is directed along the orthorhombic $b$-axis, $\langle {\bm \mu}_{\bf q} \rangle \propto (1,-1)$. If one choses the momentum transfer ${\bf q}$ along this direction, as has been done in the experiment~\cite{matsuda02}, Eq.~(\ref{eq:hint}) shows that in this case, the commensurate magnetization does not contribute to the scattering, which allows one to observe the small incommensurate peaks. This also clearly indicates that the incommensurate peaks are due to a noncollinear spin structure, $T_{\bf q} \propto {\bm \mu}^{(N)} \cdot {\bm \mu}^\perp_{\bf q}$, inconsistent with any collinear spin stripe picture.

The doping dependence of the DM spin-wave gap in the N\'eel state has been measured quite recently~\cite{gozar04}. The observed reduction of the gap is clearly due to the loss of N\'eel order, which is completely destroyed at 2\% doping.

Another remarkable feature of LSCO is the doping dependence of the uniform magnetic susceptibility in the N\'eel state at zero temperature. According to Ref.~\onlinecite{lavrov01}, the longitudinal susceptibility $\chi_b$ changes tremendously already at $x=0.01$, while the transverse susceptibilities $\chi_a$ and $\chi_c$ remain practically unchanged~\cite{comment1} compared to the undoped compound.

In the present work, based on the spiral solution of the extended $t$-$J$ model~\cite{sushkov04,sushkov05}, we explain and calculate the magnetic properties described above in  the lightly doped N\'eel state and show why the incommensurate structure in the insulating state (in both the N\'eel and the spin-glass phase) is directed along the orthorhombic $b$-direction. 

The idea of spin spiral formation in an antiferromagnet with itinerant particles goes back to Nagaoka~\cite{nagaoka66} who noted that for a sufficiently small superexchange a mobile particle turns the antiferromagnet into a ferromagnet. In relation to the $t$-$J$ model, this idea was first formulated
by Shraiman and Siggia~\cite{shraiman89}, who pointed out that for an appreciable superexchange, it is energetically favorable to allow the collinear N\'eel state to relax and form a spiral, in which holes can 
hop more easily. For a long time, the issue of stability of the spiral state remained rather controversial,
because of the negative compressibility of the uniformly doped state~\cite{kane90,auer91,singh91,igarashi92,kampf92,chubukov95}. It has been demonstrated recently~\cite{sushkov04} that the next-nearest neighbor hopping matrix elements $t'$ and $t''$ are crucially important for the stability of the uniformly doped state and that the spiral is indeed stable for physical values of $t$ and $t'$. In the uniformly doped state, the spiral is always directed along the crystal axes of the square lattice. The possibility of spiral ordering in the insulating spin-glass phase of LSCO has been pointed out in Ref.~\onlinecite{hasselmann04}. In the insulating phase, $x < 0.055$, the compressibility issue is not important because of the trapping of holes by Sr ions. Since trapped holes induce a spiral directed along the diagonal of the square lattice~\cite{sushkov05}, the direction of the incommensurate structure is rotated by $45^\circ$ at the point of the percolation-like insulator-superconductor transition. In addition, the anisotropy of the dc-conductivity in the spin-glass phase~\cite{ando02} has been explained and calculated in Ref.~\onlinecite{kotov05}.

Our paper is organized as follows. We first recall the main results of the extended $t$-$J$ model concerning the single-hole dispersion and the Sr-hole bound state in Sec.~\ref{sec:pinning} and explain why the incommensurate spin structure (spin spiral) is directed along the orthorhombic $b$-direction. In Sec.~\ref{sec:model}, we introduce the effective low-energy Hamiltonian for doped antiferromagnets within the framework of the non-linear $\sigma$-model (NLSM), which is a very convenient technical tool to deal with spin degrees of freedom in the $t$-$t'$-$t''$-$J$ model, especially when taking into account the magnetic anisotropies due to the DM and XY interactions. Numerical simulations of this model, presented in Sec.~\ref{sec:correlations}, allow us to study long-range correlations in the ground state at zero temperature. A careful comparison between our calculations of the incommensurate neutron scattering peaks and recent experimental results is contained in Sec.~\ref{sec:neutronscattering}. The evolution of the DM induced spin-wave gap upon doping is presented in Sec.~\ref{sec:gap}. Sec.~\ref{sec:susceptibilities} is then devoted to the doping-dependence of the uniform magnetic susceptibilities and finally, we present our conclusion in Sec.~\ref{sec:conclusion}.

\section{The S\lowercase{r}-hole bound state (``impurity'') and pinning of the spiral direction  to the orthorhombic $b$-axis\label{sec:pinning}}
\begin{figure}
\includegraphics[width=0.45\textwidth,clip]{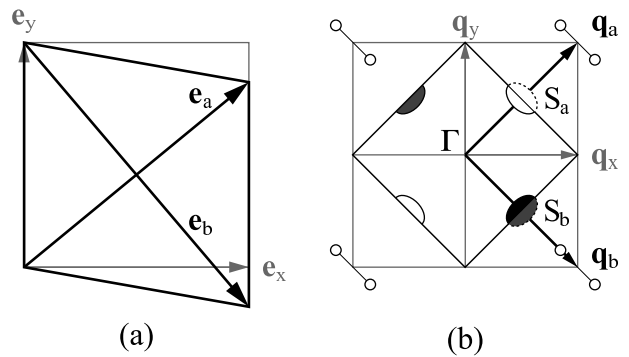}
\caption{\emph{(a) Schematic drawings of a CuO$_2$ plane in the tetragonal and orthorhombic phase. (b) Reciprocal lattice with tetragonal and orthorhombic (for simplicity shown without distortion) unit vectors. Due to the orthorhombic distortion, holes around $S_b= \left(\frac{\pi}{2},-\frac{\pi}{2}\right)$ have lower energy and the spirals induced by these holes are directed along  the orthorhombic $b$-direction. This leads to the incommensurate magnetic peaks shown as small open circles.} 
\label{fig:lattice}}
\end{figure}

Over a decade ago, the 2D $t$-$J$ model has been suggested to describe the essential low-energy physics of high-T$_c$ cuprates~\cite{anderson87,emery87,zhang88}. In its extended version, this model includes additional hopping matrix elements $t'$ and $t''$ to next-nearest neighbors. The Hamiltonian of the model is well known, see e.g. Ref.~\onlinecite{sushkov04}, and we do not present it here.  The numerical values of the parameters of the  $t$-$t'$-$t''$-$J$ model corresponding to LSCO
follow from Raman spectroscopy~\cite{tokura90} and ab-initio calculations~\cite{andersen95}. It is convenient to measure all energies in units of $J$, i.e., we set $J=125 \ meV\rightarrow1$ and  obtain $t=3.1$, $t'=-0.5$ and $t''=0.3$.  At zero doping (no holes), the extended $t$-$J$ model is equivalent to the Heisenberg model and describes the Mott insulator LCO. Removal of a single electron from this Mott insulator, or in other words injection of a hole, allows the charge carrier to propagate. Single-hole properties of the $t$-$J$ model are well understood~\cite{dagotto94}. The main features are a very flat dispersion along the edges of the magnetic Brillouin zone (MBZ) with four degenerate half-pockets centered at $S=\left(\pm\frac{\pi}{2},\pm\frac{\pi}{2}\right)$. The quasi-particle residue at the minimum of the dispersion is $Z\approx0.3$. In the full-pocket description, where two half-pockets are shifted by the AF vector ${\bf Q}_{AF}$, the two minima are located at $S_a=\left(\frac{\pi}{2},\frac{\pi}{2}\right)$ and $S_b=\left(\frac{\pi}{2},-\frac{\pi}{2}\right)$, see Fig.~\ref{fig:lattice}(b). The system is thus similar to a two-valley semiconductor.

Let us now consider a single hole trapped by the Coulomb potential of the Sr ion and refer to this bound state as the ``impurity''. Since such an ``impurity'' is an intrinsic part of LSCO, the word could be misleading and we therefore use quotation marks to avoid confusion. An ``impurity'' has a hydrogen-like structure.
In coordinate space, one part of the hole wave-function corresponds to the $1s$ bound state and depends smoothly on $r$, while a second part, dictated by the Bloch theorem, is rapidly varying with $r$ and different for holes located near $S_a$ or $S_b$. A pseudo-spin finally indicates if the hole sits on the $\uparrow$- or $\downarrow$-sublattice. The ground state can therefore be described by two quantum numbers, the hole-pocket and the pseudo-spin associated with the sublattice. In this case, the bound state in the tetragonal phase is four-fold degenerate $\left\{1s_{a\uparrow}, 1s_{a\downarrow}, 1s_{b\uparrow}, 1s_{b\downarrow}\right\}$. However, in the low-temperature orthorhombic phase of LSCO we are interested in, the orthorhombic $b$-direction is slightly longer than the $a$-direction, as illustrated in Fig.~\ref{fig:lattice}(a), $a=5.349$~\AA, and $b=5.430$~\AA, see e.g., Ref.~\onlinecite{kastner98}. To first order, this distortion only influences the diagonal hopping matrix element $t'$, whose contribution to the dispersion is equal to
\begin{equation*}
\epsilon_{\bf q}=Z t'_a\left(e^{iq_x+iq_y}+\text{H.c.}\right)+Z t'_b\left(e^{-iq_x+iq_y}+\text{H.c.}\right) \ .
\end{equation*}
Using the estimations $t'_a = t' \propto 1/a$ and $t'_b \propto 1/b$, it is easy to see that the degeneracy associated with the pockets is lifted and holes in the vicinity of $S_b$ have slightly lower energy
\begin{equation} \label{eq:deltaepsilon}
\Delta\epsilon=\epsilon_{S_a}-\epsilon_{S_b} =4Z(t'_b-t'_a) \sim 4 Z t' (a/b-1) \sim 1 \ meV \ .
\end{equation}
The bound state is thus only doubly degenerate $\left\{1s_{b\uparrow},  1s_{b\downarrow}\right\}$. The spiral formed by the trapped hole (``impurity'') is orthogonal to the corresponding face of the MBZ~\cite{sushkov05}. Hence this immediately explains, why in the orthorhombic phase spirals are directed along the $b$-direction. This consideration is equally applicable to both the N\'eel and the spin-glass phase. The spiral pinning energy is given by $E_{pin}\sim \xi^2 \, x \Delta\epsilon$, $x$ being the hole concentration and $\xi$ the correlation length in units of lattice spacings. The correlation length is anisotropic and temperature dependent, and due to frustration it remains finite, even at zero temperature, see Ref.~\onlinecite{sushkov05}. Using the experimental value~\cite{matsuda00b} at $x \sim 0.02$ obtained at $T=1.6 K$, $\xi^2 \sim 138$, one finds $E_{pin} \sim 3 \ meV \sim 35 K$. We would like to emphasize that Eq.~(\ref{eq:deltaepsilon}) only provides a crude estimate, an accurate LDA calculation of $t'_b$ and $t'_a$ is necessary to determine the precise value of $\Delta\epsilon$.

\section{Effective low energy Hamiltonian of the doped antiferromagnet. The NLSM approach\label{sec:model}}
The NLSM is a very convenient tool to describe the low energy dynamics of the weakly doped $t$-$J$ model, representing underdoped cuprates. In this framework, the staggered component of the copper spins located in a {\it single layer} of LSCO is represented by a continuous vector field ${\vec n}\left({\bf r}\right)$ of unit length ${\vec n}\left({\bf r}\right)^2=1$. To avoid confusion, we denote vectors acting in the three-dimensional (3D) spin space by arrows and vectors acting in the 2D coordinate space by the usual bold font. Throughout this paper, we adopt the orthorhombic coordinate system shown in Fig.~\ref{fig:lattice}, with unit vectors ${\vec e}_{\alpha}$, $\alpha=a,b,c$. Coordinate and spin space are linked through the pinning of the  commensurate N\'eel magnetization to the orthorhombic $b$-axis. It is therefore convenient to use the same coordinate system in both cases with real space unit vectors ${\bf e}_{\alpha}={\vec e}_{\alpha}$.

An elegant way to incorporate the DM and XY anisotropies in the NSLM, crucially important for an accurate description of the N\'eel state of LSCO, has been proposed quite recently by Silva Neto {\it et al.}. In the static limit, the energy of the spin system reads~\cite{silvaneto05}
\begin{multline} \label{eq:en}
E_n=\frac{\chi_\perp}{2} \int d^2r \left\{c^2 \left({\bf \nabla}{\vec n}\left({\bf r}\right)\right)^2 \right. \\+ \left. \left({\vec D}\cdot{\vec n}\left({\bf r}\right)\right)^2 +\Gamma_c n_c^2\left({\bf r}\right) \right\} \ .
\end{multline}
Here $\chi_{\perp}\approx 0.065$ is the magnetic susceptibility of the Heisenberg model, $c=\sqrt{\rho_s/\chi_\perp} \approx 1.66$ the spin-wave velocity and $\rho_s \approx 0.18$ is the spin stiffness. The anisotropies are due to the DM interaction, with a DM vector directed along the orthorhombic $a$-direction ${\vec D}\approx0.02 {\vec e}_{a}$, and the XY-term, which leads to $\sqrt{\Gamma_c} \approx0.04$.

The trapped hole has a hydrogen-like ground state wave-function
\begin{equation*}
\psi\left({\bf r}\right) = \Psi \chi \left({\bf r}\right) = \Psi \sqrt{\frac{2}{\pi}} \kappa e^{-\kappa r} \ ,
\end{equation*}
where $\Psi$ is a two-component spinor (independent of ${\bf r}$) describing the pseudo-spin. Note that according to the previous section the hole resides in the $S_b$ hole-pocket, however, we do not write this index explicitly in the wave function. The value of $\kappa$ slightly depends on doping. It decreases with increasing doping, because of the Coulomb screening. According to data on hopping conductivity~\cite{chen95} $\kappa \approx 0.3 - 0.4$ at $x=0.002$ and very recent preliminary data taken at $x=0.03$ indicate~\cite{panagopoulos05} that $\kappa \approx 0.2$. The hole energy inside the ``impurity'' can be written as
\begin{equation} \label{eq:epsi}
E_{\psi}=\int d^2r \ \psi^{\dag}\left({\bf r}\right) \left(-\beta\frac{{\bf \nabla}^2}{2}-\frac{e^2}{\epsilon_e r}\right)\psi\left({\bf r}\right) \  ,
\end{equation}
where $\beta \approx 2$ is the inverse mass of the trapped hole~\cite{sushkov04}, and $\epsilon_e$ is the effective dielectric constant. Strictly speaking, one has to write the Coulomb energy as $e^2/(\epsilon_e\sqrt{d^2+r^2})$, where $d\approx 2$~\AA is the distance from the CuO$_2$ plane to the Sr ion, but since $r \gg d$ one can safely neglect the $d^2$ term.

The interaction energy of a hole with the spin degrees of freedom reads~\cite{sushkov05}
\begin{equation} \label{eq:epsin}
E_{\psi n}= \sqrt{2}g\int d^2r \ \left[{\vec n}\left({\bf r}\right) \times\psi^{\dag}\left({\bf r}\right) {\vec \sigma} \psi\left({\bf r}\right) \right]\left({\bf e}\cdot{\bf \nabla}\right){\vec n}\left({\bf r}\right) \ .
\end{equation}
Here ${\bf e}$ is a unit vector orthogonal to the face of the MBZ, ${\bf e}={\bf e}_b$, since the hole resides in the $S_b$ hole-pocket and $\vec \sigma$ are the Pauli matrices. The coupling constant $g\approx Z t\approx 1$ has been calculated previously~\cite{igarashi92,sushkov04} within the $t$-$t'$-$t''$-$J$ model.

The restriction to the lightly doped N\'eel phase allows us to expand the ${\vec n}$-field around the dominating commensurate order directed along the $b$-axis
\begin{equation} \label{eq:n}
{\vec n}\left(\bf r\right) = {\vec e}_{b}+{\vec \pi}\left(\bf r\right) = {\vec e}_{b}+ \sum_{\alpha=a,c} \pi^\alpha \left({\bf r}\right) {\vec e}_{\alpha} \ ,
\end{equation}
where ${\vec \pi}$ is a small vector field orthogonal to the $b$-direction, i.~e. ${\vec \pi}^{\,2}\ll1$.
Using this approximation, the total energy of a system with $N$ ``impurities'' can be expressed as
\begin{align} \label{eq:etot}
E &=E_n+E_\Psi+E_{\Psi n} \\
&= \int d^2 r \left\{ \frac{\rho_s}{2} \sum_{\alpha=a,c} \left\{ \left[ {\bf \nabla} \pi^\alpha \left({\bf r}\right) \right]^2 + M_\alpha^2 \left[ \pi^\alpha\left({\bf r}\right)\right]^2 \right\} \right. \nonumber \\
& \left. \quad + \sqrt{2} g \sum_{i=1}^N \rho \left({\bf r}-{\bf r}_i\right)   {\vec m}_i \left({\bf e} \cdot {\bf \nabla}\right) {\vec \pi}\left({\bf r}\right)\right\}+ E_0\nonumber \ ,
\end{align}
where ${\vec m}_i={\vec e}_{b} \times \left( \Psi_i^\dag {\vec \sigma} \Psi_i\right)$ are the 
directors of the ``impurities'', $\left|{\vec m}_i\right|=1$, $\rho\left({\bf r}\right)=\chi^2\left({\bf r}\right)$, and 
$E_0=N\left(\frac{\beta \kappa^2}{2}-2\kappa e^2\right)$. Since $E_0$ is independent of ${\vec m}_i$ and ${\vec \pi}$, it only leads to a shift in energy which can be safely omitted for the purpose of the present work. The ``masses'' follow from~(\ref{eq:en}): 
\begin{equation*}
M_a= D/c=0.012 \quad\text{and}\quad M_c= \sqrt{\Gamma_c}/c=0.024 \ .
\end{equation*}
Note that this definition of the ``masses'' differs from the standard one by a factor $1/c$.

\subsection{Solution for an isolated ``impurity''}
Let us for clarity reasons introduce a special notation for the single ``impurity'' problem and 
use the vector field ${\vec \theta}$ instead of ${\vec \pi}$ in this case, i.e., analogous to the definition~(\ref{eq:n}), we use
\begin{equation*}
{\vec n}\left(\bf r\right) = {\vec e}_{b}+{\vec \theta}\left(\bf r\right) = {\vec e}_{b} + \sum_{\alpha=a,c} \theta^\alpha \left({\bf r}\right) {\vec e}_{\alpha} \ ,
\end{equation*}
and assume ${\vec \theta}^{\,2} \ll 1$. According to Ref.~\onlinecite{sushkov05}, the single
``impurity'' solution which minimizes the total energy~(\ref{eq:etot}) satisfies the equation
\begin{equation} \label{eq:eqtheta}
\left(-\triangle + M_{\alpha}^2\right) \theta^\alpha\left(\bf r\right)  - m^\alpha {\cal M} ({\bf e}\cdot{\bf \nabla})\rho\left(\bf r\right)  = 0\ .
\end{equation} 
Here $m^\alpha$ is the $\alpha$-component of the director ${\vec m}$, and ${\cal M} =\sqrt{2}g/\rho_s \approx 8$ is the effective dipole moment of the ``impurity''.

The solution of Eq.~(\ref{eq:eqtheta}) in coordinate and  momentum representation reads
\begin{align} \label{eq:theta}
\theta^\alpha \left({\bf r}\right) &= m^\alpha \frac{{\cal M}}{2\pi} \frac{{\bf e}\cdot{\bf r}}{r^2} \left\{\left(1+2 \kappa r\right) e^{-2 \kappa r}- M_\alpha r K_1\left( M_\alpha r\right)\right\} \ ,  \nonumber \\
\theta^\alpha_{\bf q} &= m^\alpha {\cal M} \frac{i\,{\bf e}\cdot{\bf q}} {q^2+ M_{\alpha}^2} \rho_{\bf q} \ ,
\end{align}
where $K_1$ is a modified Bessel function of the second kind and we make use of the fact that $M_\alpha \ll \kappa$.  The Fourier transformation $\rho_{\bf q}$ of $\rho\left( {\bf r}\right)$ is given by
\begin{equation} \label{eq:rho}
\rho_{\bf q} = \frac{8 \kappa^3}{\left(4\kappa^2+q^2\right)^{3/2}} \approx 1 \quad \left(q \ll \kappa\right) \ .
\end{equation}
 The above solutions are valid in the case of an isolated copper-oxide plane.
 
\subsubsection{Influence of neighboring CuO$_2$ planes}
Let us consider the modification of the solution~(\ref{eq:theta}) due to the interaction with other copper-oxide planes. The energy spectrum of spin waves in LCO reads~\cite{keimer92}
\begin{align*}
\omega\left(q_x,q_y,q_z\right) &= 2.32\,J\sqrt{(1+\alpha_\perp/2)^2-(\gamma_{\parallel}+\alpha_\perp\gamma_\perp/2)^2} \\
\gamma_\parallel &=\frac{1}{2}(\cos{q_x}+\cos{q_y})\, ,\; \gamma_\perp= \cos{q_z}\, .
\end{align*} 
Here $\alpha_{\perp}\approx 5\cdot10^{-5}$ describes the superexchange between the planes. We disregard DM and XY anisotropies, because they are already taken into account in the effective NLSM. Incorporating the $q_z$ dependence of this spectrum in Eq.~(\ref{eq:theta}) in the limit where ${\bf q}=(q_x,q_y)$ is small, we get
\begin{equation} \label{eq:theta3d}
\tilde \theta^{\alpha}\left({\bf q},q_z\right)\propto \frac{1}{q_x^2+q_y^2+4\alpha_\perp\sin^2(q_z/2)+M_\alpha^2} \, .
\end{equation}
In order to find an effective expression of~(\ref{eq:theta3d}) in the single-layer approximation used to derive Eq.~(\ref{eq:theta}), we integrate~(\ref{eq:theta3d}) over all momenta $q_z$ and obtain 
\begin{multline*}
\theta^{\alpha}_{\bf q} \propto \frac{1}{\pi} \int_0^\pi  \, \frac{dq_z}{q_x^2+q_y^2+4\alpha_\perp\sin^2(q_z/2)+M_\alpha^2} \\
=\frac{1}{\sqrt{\left(q_x^2+q_y^2+M_\alpha^2\right)\left(4\alpha_\perp+q_x^2+q_y^2+M_\alpha^2\right)}} \ ,
\end{multline*} 
which can be approximated by the original expression~(\ref{eq:theta}) with the substitution of an effective mass
\begin{equation*}
M_\alpha \rightarrow M_\alpha^\text{eff}=M_\alpha\left(1+\frac{4 \alpha_\perp}{M_\alpha^2}\right)^{1/4} \ ,
\end{equation*}
i.e. $M_a^\text{eff}=0.015$ and $M_c^\text{eff}=0.026$. We use these values for further calculations.

\subsection{Solution for multiple ``impurities''}
It is straightforward to generalize the single ``impurity'' solution~(\ref{eq:theta}) to a  system consisting of $N$ ``impurities''. Because of the assumption ${\vec \pi}^{\,2} \ll1$, the ${\vec \pi}$-field is the linear superposition of $N$ independent local spirals
\begin{equation} \label{eq:pi}
{\vec \pi}\left({\bf r}\right) = \sum_{\alpha=a,c} \sum_{i=1}^N \theta_i^\alpha \left({\bf r}-{\bf r}_i\right) {\vec e}_{\alpha}\ ,
\end{equation}
where $\theta^\alpha \left({\bf r}\right)$ is the single ``impurity'' solution given by 
Eq.~(\ref{eq:theta}). 

As already pointed out in Ref.~\onlinecite{sushkov05}, the single ``impurity'' solution is degenerate with respect to the orientation of the director ${\vec m}$, since minimizing the total energy~(\ref{eq:etot}) only enforces the ${\vec \theta}$-field to be parallel to ${\vec m}$. This degeneracy is lifted by the interaction with other ``impurities''. Having $N$ holes trapped by $N$ Sr ions, one can calculate the effective interaction energy between ``impurities'' due to perturbation of  the NLSM in the limit where the distance $r$ between two ``impurities'' is sufficiently large, $r=\left| {\bf r}_i-{\bf r}_j \right| > 1/\kappa \sim 3$. Let us substitute~(\ref{eq:pi}) in the total energy~(\ref{eq:etot}) and split the result into an effective interaction energy $U$ (containing contributions $i\ne j$) and a self-energy $\Sigma$ (arising from contributions $i=j$). Omitting terms independent of ${\vec m}$, which just shift the energy, we find
\begin{equation} \label{eq:etot2}
E=U+\Sigma=\sum_{\alpha=a,c} \sum_{i \ne j}^N U^\alpha_{i,j} + \sum_{\alpha=a,c} \sum_i^N \Sigma^\alpha_i
\end{equation}
where
\begin{align} \label{eq:uij}
U_{ij}^\alpha &=\frac{\rho_s{\cal M}^2}{4\pi} M_{\alpha}^2 m^\alpha_i m^\alpha_j \bigg\{ ({\bf e}_i \cdot{\bf e}_j) K_0(M_{\alpha}r) \nonumber \\
&\quad -\left[({\bf e}_i\cdot{\bf e}_j)- \frac{2({\bf e}_i\cdot{\bf r})({\bf e}_j\cdot{\bf r})}{r^2}\right] K_2(M_{\alpha} r )\bigg\} \ ,
\end{align}
with the modified Bessel functions $K_n$, and
\begin{align} \label{eq:ei}
\Sigma^\alpha_i = \frac{\rho_s{\cal M}^2}{8\pi} M_\alpha^2 \left( m^\alpha_i\right)^2 \ln\left(\frac{\kappa}{M_\alpha}\right) \ .
\end{align}
For $M_{\alpha} \rightarrow 0$, Eq.~(\ref{eq:uij}) agrees with Refs.~\onlinecite{glazman90,cherepanov99,hasselmann04}. We would like to stress that even though the interaction~(\ref{eq:uij}) looks similar to the usual electrostatic dipole-dipole interaction, it is substantially different: Firstly,  the dipole director ${\vec m}_i$ which determines the {\it polarization} of the spiral is decoupled from the vector ${\bf e}_i$ which determines the {\it direction} of the spiral. Secondly, the sign of~(\ref{eq:uij}) is opposite to what one naively expects from the analogy to electrostatics and thirdly, the interaction is of finite range due to nonzero masses. Although the self-energy~(\ref{eq:ei}) is proportional to $M_\alpha^2$ and therefore small numerically, our results presented in Sec.~\ref{sec:correlations} show that its contribution is crucially important and leads to an alignment of the ``impurity'' directors ${\vec m}_i$ along the $a$-direction, because $M_a < M_c$.

\subsubsection{``Molecular impurities''}
The effective interaction~(\ref{eq:uij}) is valid for ``impurities'' separated by a distance $r > 1/\kappa$. In this case, the orthorhombic distortion favors holes to reside in the vicinity of $S_b$, and the vectors ${\bf e}_i$ in~(\ref{eq:uij}) are all directed along the $b$-axis, i.e., ${\bf e}_i = {\bf e}_b$. However, as soon as two ``impurities'' are sitting very close to each other, the holes can no longer be treated as independent particles but have to obey the Pauli exclusion principle, as  explained in Ref.~\onlinecite{sushkov05}. In order to elucidate this mechanism in more detail, it is convenient to distinguish between ``atomic'' and ``molecular impurities'': an ``atomic impurity'' is just a redefinition of the bound state formed by a Sr ion and the trapped hole (in analogy with the hydrogen atom), whereas a ``molecule'' describes two or more ``atomic impurities'' with noticeably overlapping hole wave functions. The  formation of ``molecules'' is entirely due to the Pauli blocking: For two or more well separated ``atomic impurities'', the Pauli exclusion principle does not apply and the orthorhombic distortion favors holes residing in the pocket centered at $S_b$. But as soon as ``atomic impurity'' wave-functions have a non-negligible overlap, the Pauli blocking sets in and two holes in the same hole-pocket must have opposite pseudo-spins, which prevents the formation of a local spiral and therefore does not lead to a gain in energy. In this situation, it will be energetically favorable to place the holes in different pockets, which, due to the orthorhombic distortion costs an energy $\Delta\epsilon\approx 1 \ meV$~(\ref{eq:deltaepsilon}) but allows the formation of a local spiral~\cite{sushkov05}. For parallel alignment of the ``impurity'' directors this gain in energy is of about $3-5\ meV$. A ``molecule'' can therefore be represented by the director ${\vec m}$ and the vector ${\bf e}={\bf e}_a \pm {\bf e}_b$, since it is a superposition of two holes in pockets $S_a$ and $S_b$ with two times the self-energy~(\ref{eq:ei}) of an ``atomic impurity''. The critical distance between ``atomic impurities'', below which ``molecules'' are formed, is not well defined. Clearly, it is smaller than the average distance between ``impurities'' at the insulator-superconductor transition (percolation)  at doping $x=0.055$, i.e. $r_c < 1/\sqrt{x} \approx 4.3$ lattice spacings. Although the formation of ``molecules'' is crucially important to explain the jump of the incommensurability direction at the insulator-superconductor transition~\cite{sushkov05}, it is negligible in the lightly doped N\'eel phase considered in this work. For instance, for $r_c=2$ at doping $x=0.01$ only 6\% of the ``impurities'' form ``molecules'' - a contribution which can be safely neglected. This conclusion is supported by extensive numerical simulations for $r_c \leq 3$ lattice spacings, which do not indicate any modifications due to ``molecules''.

\section{Ground state and correlation functions\label{sec:correlations}}
Let us now investigate the properties of the zero-temperature ground state of the  system. The basic assumption behind the derivation of the effective  interaction~(\ref{eq:etot2}) is the existence of a dominating commensurate N\'eel  order~(\ref{eq:n}), which restricts our analysis to the low-doping situation $x \ll 0.02$. From a technical point of view, the Hamiltonian~(\ref{eq:etot2}) represents a set of
interacting dipoles ${\vec m}_i$. In the static limit, where any kinetic energy terms are absent, we are dealing with a classical problem. This classical approximation is justified by the large value of the effective dipole moment ${\cal M} \approx 8$ and by the long-range character of the interaction. In order to describe realistic experimental situations, with samples of LSCO  consisting of many copper-oxide layers, each of them containing a random arrangement of  ``impurities'' and all of them contributing to the measurement, we have to average the  quantities we calculate over many different realizations of random dipole positions. To find the ground state of the Hamiltonian~(\ref{eq:etot2}), we perform classical zero-temperature Monte-Carlo simulations with up to $N=200$ randomly distributed dipoles on a square lattice. These dipoles are separated by an average distance $l=1/\sqrt{x}$ lattice spacings. In accordance with the previous discussion of the spiral direction pinning, we set ${\bf e}_i ={\bf e}_b$. Our results clearly indicate that the self-energy term~(\ref{eq:ei}) leads to a pinning of the  dipoles  along the orthorhombic $a$-direction, ${\vec m}_i \propto \pm {\vec e}_a$, because of the smaller mass. This effect allows  us to divide the task of finding the ground state into two parts: in a first step, we only  consider the Ising-like situation, where all dipoles are aligned along the $a$-axis. Starting  from random initial conditions, we find the ground state of a given realization by exactly minimizing  clusters of eight dipoles at a time. Our algorithm generates random walks through the system and forms the clusters at a given site according to the strength of the interaction~(\ref{eq:uij}) with neighboring dipoles. The total energy is then minimized with respect to the dipoles in this cluster and the algorithm proceeds to the next site of the walk. In a second step, we perturbate this collinear arrangement  and allow the dipoles to relax and have a nonzero component along the $c$-direction. We find that for a given number of realizations, the percentage of ground states with noncollinear dipole alignment decreases with increasing system sizes. However, this observation could very  well be influenced by the difficulty of finding the optimal noncollinear dipole arrangement in larger systems.

\begin{figure}
\includegraphics[width=0.45\textwidth,clip]{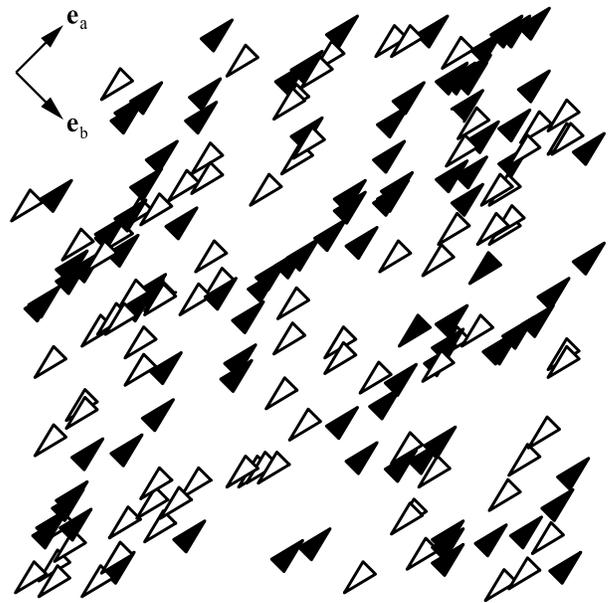}
\caption{\emph{Example of a ground state dipole arrangement for 1\% doping. The dipoles form clusters stretched along the orthorhombic $a$-direction.} \label{fig:vectorfield1}}
\end{figure}

\begin{figure*}
\includegraphics[width=0.9\textwidth,clip]{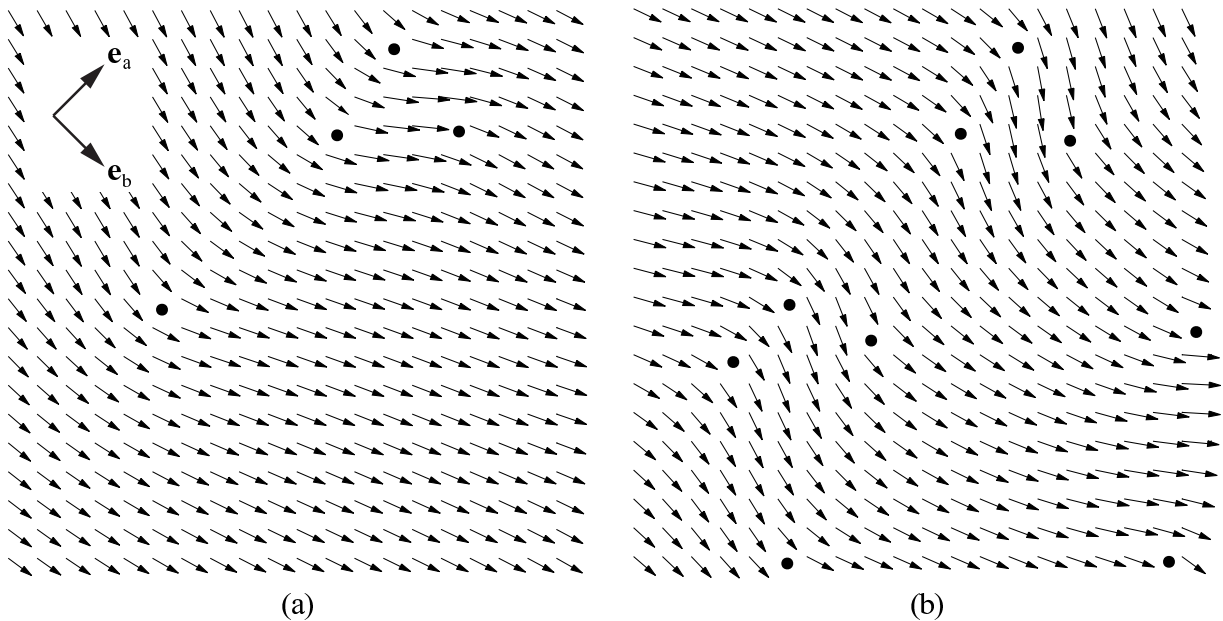}
\caption{\emph{${\vec n}$-fields derived from two realizations of random impurity distributions at dopings $x=0.005$ (a) and $x=0.014$ (b). Both systems consist of $L=60\times60$ sites and contain $N=18$ and $N=50$ dipoles, respectively. For readability, only a characteristic part of these systems is shown, with circles denoting the positions of ``impurities''. Spin and coordinate space are linked by the pinning of the commensurate magnetization along the orthorhombic $b$-direction. Clearly, doped holes lead to a destruction of N\'eel order, experimentally observed at $x=0.02$.} \label{fig:vectorfield2}}
\end{figure*}

A typical ground state configuration for a given realization of random dipole positions at $x=0.01$ is shown in Fig.~\ref{fig:vectorfield1}. 
One clearly identifies domains with parallel dipole alignments. The size of these domains along the $a$-direction is substantially larger than along the $b$-direction. This picture also reflects the ferromagnetic (resp. antiferromagnetic) character of  the dipole-dipole interaction~(\ref{eq:uij}) in the $a$-direction (resp. $b$-direction). Having found the ground state of a given realization in terms of the ``impurity'' directors ${\vec m}$, one can easily calculate the resulting ${\vec n}$-field, using  Eqs.~(\ref{eq:n}), (\ref{eq:pi}) and (\ref{eq:theta}). As an example,  Figs.~\ref{fig:vectorfield2}(a) and (b) show the ${\vec n}$-field derived from the ground  state dipole arrangement for dopings $x=0.005$ and $x=0.014$, respectively.  For readability, we only show a small part of these systems, where ``impurities'' are represented by filled circles. 
One can see that the initially dominating orientation along the $b$-direction becomes weaker with increasing doping.

\begin{figure}
\includegraphics[width=0.45\textwidth,clip]{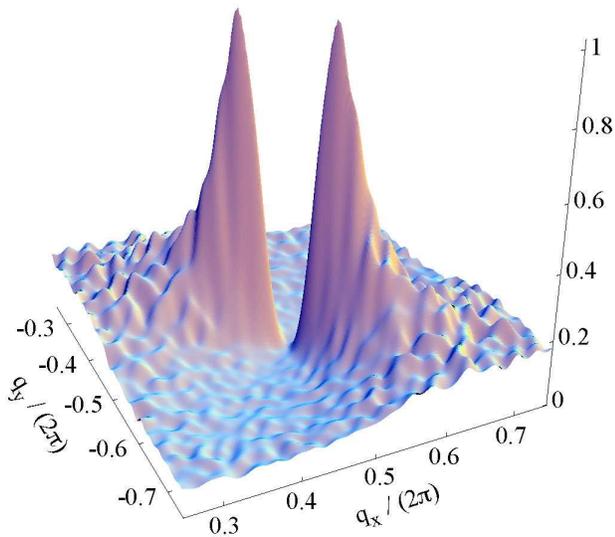}
\caption{\emph{(Color online). Normalized correlation function $D^a_{\bf q}$ obtained from ground states of model~(\ref{eq:etot2}) for systems with $L=141\times141$ sites and $N=200$ ($x=0.01$) ``impurities''. The statistical average is taken over 200 realizations. Due to the pinning along the $a$-direction, $D^c_{\bf q}=0$.} \label{fig:correlations1}}
\end{figure}

\begin{figure}
\includegraphics[width=0.45\textwidth,clip]{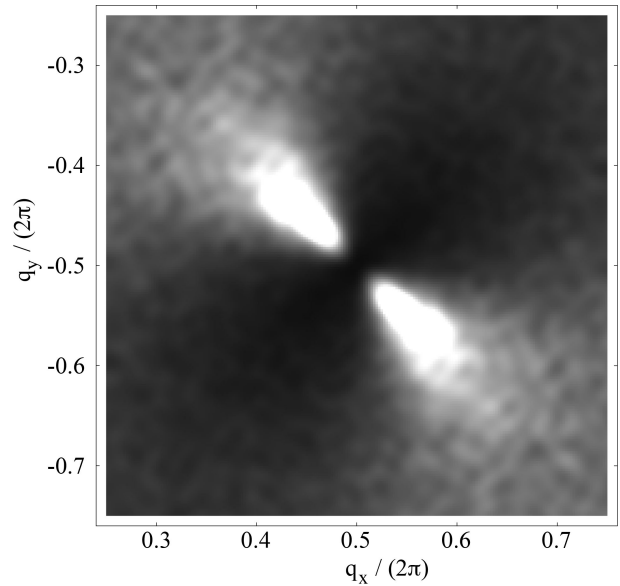}
\caption{\emph{Density plot of $D^a_{\bf q}$ shown in Fig.~\ref{fig:correlations1}. The plot clearly reveals the asymmetry of the incommensurate peaks.  The width along the $b$-direction is around twice as large as along the $a$-axis.} \label{fig:correlations2}}
\end{figure}

\begin{figure}
\includegraphics[width=0.45\textwidth,clip]{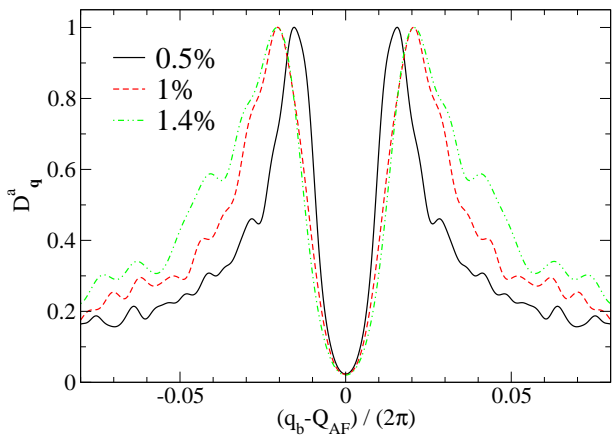}
\caption{\emph{(Color online). Slices of the normalized correlator $D^a_{\bf q}$ along the $b$-direction for three different dopings. The widths of the peaks clearly increases with increasing doping.} \label{fig:correlations3}}
\end{figure}

In order to characterize the ground state, we define a correlation function, closely related to the neutron scattering cross section, see Eq.~(\ref{eq:crosssection}) in the following Sec. 
\begin{equation} \label{eq:d}
D^\alpha_{\bf q}=\frac{1}{N} \sum_{i,j} m^\alpha_i m^\alpha_j e^{i{\bf q}\cdot \left({\bf r}_i-{\bf r}_j\right)} \ .\end{equation}
Due to the very strong pinning of the dipoles along the $a$-direction, their $c$-components are negligible and the corresponding correlations vanish, i.e., $D^c \left({\bf q}\right)=0$. The asymptotic behavior is thus given by $D^a \left({\bf q}\right) \rightarrow1$ for $q \rightarrow \infty$.
The common feature of $D^a_{\bf q}$, obtained for different hole concentrations $x$ is a pronounced peak centered on the orthorhombic $b$-axis, as shown in Fig.~\ref{fig:correlations1} for $x=0.01$. The statistical average of $D^a_{\bf q}$ is obtained from 200 realizations of random dipole distributions in systems with $L=141\times141$ sites and $N=200$ ``impurities''.
Statistical fluctuations are negligible compared to these well pronounced incommensurate peaks, which are manifestations of the dipole domains shown in Fig.~\ref{fig:vectorfield1}. A density plot of $D^a \left({\bf q}\right)$ (same data as Fig.~\ref{fig:correlations1}) shown in  Fig.~\ref{fig:correlations2} reveals the anisotropy of the peaks.
The width along the  $b$-direction is around twice as large as along the $a$-direction. Most likely, we slightly overestimate the width in the $b$-direction and substantially overestimate the width in the $a$-direction due to finite-size effects. The substantial underestimation of the correlation length (inversly proportional to the width of the peak) along the $a$-direction is clearly illustrated in Fig.~\ref{fig:vectorfield1}. The size of the domain along the $a$-direction is comparable to the size of the lattice. Unfortunately, the convergence of our Monte-Carlo procedure is getting very slow when simulating larger systems, so that we are limited to this size. In Fig.~\ref{fig:correlations3}, we show slices through the maximum of the peaks parallel to the $b$-axis. In this lightly doped regime, the correlation length $\xi_b$ clearly decreases with increasing doping.
Note that the correlations length extracted from neutron scattering experiments~\cite{matsuda02} cannot be directly compared to the dipole-dipole correlations $D_{\bf q}^\alpha$ (\ref{eq:d}), because the neutron scattering cross-section is the product of the correlator with the single-dipole contribution, see Eq.~(\ref{eq:crosssection}) in the following Sec.

\section{Incommensurate neutron scattering peaks\label{sec:neutronscattering}}
As explained in the introduction, the experimental conditions~\cite{matsuda02} are such as neutrons only interact with the projections of electron spins orthogonal to the $b$-axis, which in terms of the NLSM is just the ${\vec \pi}$-field. Using Eqs.~(\ref{eq:pi}), (\ref{eq:theta}) and (\ref{eq:d}), we find the following neutron scattering cross section
\begin{equation} \label{eq:crosssection}
\left| T_{\bf q} \right|^2 \propto
 \left| {\vec \mu}^\perp_{\bf q} \right|^2 \propto \left| {\vec \pi}_{\bf q}\right|^2  =
x \Omega \sum_{\alpha=a,c} G^\alpha_{\bf q} D^\alpha_{\bf q} \  ,
\end{equation}
where $\Omega$ is the area of the sample and $G^\alpha_{\bf q}$ is the single-dipole contribution given by
\begin{equation} \label{eq:g}
G^\alpha_{\bf q} ={\cal M}^2 \frac{\left({\bf q} \cdot {\bf e}_b\right)^2 } {\left(q^2+M_{\alpha}^2\right)^2} \rho_{\bf q}^2 \ .
\end{equation}
Experimental results are available for 1\%, 1.4\% and 1.8\% doping, shown in Fig.~3 of Ref.~\onlinecite{matsuda02} and reproduced in Fig.~\ref{fig:neutronscattering}. Since we assume the presence of a dominant N\'eel order, our results are reliable only at small doping, far from the transition to the spin-glass phase. We therefore only consider $x=0.01$ and $x=0.014$.  It is explained in Ref.~\onlinecite{matsuda02} that the small asymmetry experimentally observed for 1.4\% doping [Fig.~\ref{fig:neutronscattering}(b)] arises due to different twinning directions (there are four possible equivalent orthorhombic distortions of the tetragonal lattice) present in the crystal.
\begin{figure}
\includegraphics[width=0.45\textwidth,clip]{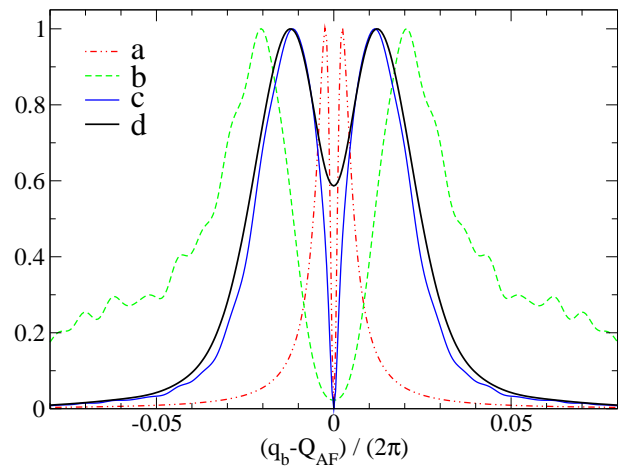}
\caption{\emph{(Color online). Normalized neutron scattering cross section along the orthorhombic $b$-axis for x=0.01. According to Eq.~(\ref{eq:crosssection}), $G^a_{\bf q}$ (a) represents the single dipole contribution and $D_{\bf q}^a$ (b) the dipole-dipole correlation function. Neither of these contributions alone is sufficient to characterize the resulting cross-section $\left| T_{\bf q} \right|^2$ (c) given by their product. For comparison with experiments, we convolute $\left| T_{\bf q} \right|^2$ with a Gaussian to take into account the finite experimental resolution (d). Note that all quantities are normalized to unity.} \label{fig:correlations4}}
\end{figure}
Eq.~(\ref{eq:crosssection}) clearly shows that the incommensurate peak observed in neutron scattering is  due to two effects, because it is the product of the single-dipole contribution $G^a_{\bf q}$ with the correlation function $D^a_{\bf q}$. In the limit of noninteracting dipoles the correlator is trivial, $D_{\bf q}^a=1$, and hence the maximum of the neutron peak, solely determined by the smaller mass, is found at $q=M_a$.  However in the case of interacting dipoles, there is a second, much broader peak in the correlation function $D^a_{\bf q}$ which has to be taken into account. Fig.~\ref{fig:correlations4} clearly shows  that neither of the two contributions alone is sufficient to characterize the resulting neutron 
scattering cross section. Only the product~(\ref{eq:crosssection}) gives the right answer.
\begin{figure}
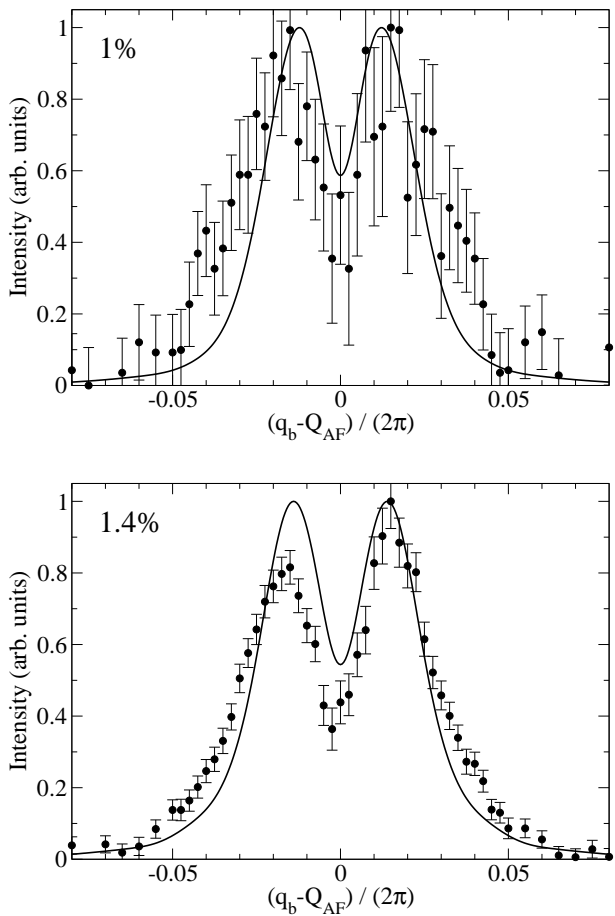

\includegraphics[width=0.45\textwidth,clip]{fig8a.eps} \\*[0.5cm]
\includegraphics[width=0.45\textwidth,clip]{fig8b.eps}
\caption{\emph{Neutron scattering probability for  $x=0.01$ and $x=0.014$. The dots correspond to experimental observations taken from Fig.~3 in Ref.~\onlinecite{matsuda02}, with normalized intensity. The curves represent our simulations convoluted with a gaussian to take into account the finite experimental resolution. The agreement between simulations (containing no fitting parameters at all) and experiments is remarkable.} \label{fig:neutronscattering}}
\end{figure}
In order to compare our results with experiments, we take into account the finite experimental resolution
by a convolution with a gaussian of half-width $\Gamma_\text{resol}=0.005$. This convolution only leads to a nonzero value at $q=0$ but does not have any influence on the positions and the widths of the peaks, as illustrated in Fig.~\ref{fig:correlations4}. A comparison with experiments, presented in Fig.~\ref{fig:neutronscattering}, shows that our simulations are in remarkable agreement with experimental observations, especially considering the fact that these curves do not contain any fitting parameters. Although the finite-size effects encountered in our simulations influence the correlator $D^a_{\bf q}$ (see Sec.~\ref{sec:correlations}), they are practically absent in the neutron scattering cross section. Simulations of systems with 140 dipoles (not shown) lead to almost identical curves in Fig.~\ref{fig:neutronscattering}.

\section{Doping dependence of the zero momentum DM spin-wave gap\label{sec:gap}}
Gozar and co-workers~\cite{gozar04} have recently measured the DM induced spin-wave gap in LSCO at zero momentum using Raman scattering. Their result for zero doping $\Delta_{DM}=D\approx 17.4cm^{-1}\approx 2.2meV$ agrees reasonably well with the value $D\approx 2.5meV$ known from  neutron scattering~\cite{keimer92b}. According to Ref.~\onlinecite{gozar04}, the gap is reduced by 28\% to $\Delta_{DM}\approx 12.5cm^{-1}$ at 1\% doping. As one should expect, the gap vanishes at $x=0.02$, where the N\'eel order is completely destroyed. Since our approach relies on the existence of an ordered state, it does not allow to describe the transition to the spin-glass phase at $x=0.02$, but for $x=0.01$ the approximation is well justified.

Since the anisotropy due to the DM interaction $D$ is independent of doping, the relation $\Delta_{DM}=D=cM_a$ is only valid at zero doping. In order to determine the spin-wave gap at finite doping, we consider an excitation with zero momentum. In the framework of the NLSM, the ground state of the Hamiltonian~(\ref{eq:etot}) can be represented as ${\vec n}=(n^a,n^b,n^c)=\left(\sin\left[\pi^a({\bf r})\right],\cos\left[\pi^a({\bf r})\right],0\right)$, with the ${\vec \pi}$-field found in the previous Secs. Since this vector-field is small, $\pi^a\ll 1$, one can expand the trigonometric functions in powers of $\pi^a$. In order to preserve rotational symmetry, we expand only the final answer and use the exact expression in our calculations. A zero momentum excitation corresponds to a global rotation of the ground state ${\vec n}$-field by an angle $\alpha \ll 1$
\begin{equation*}
{\vec n}({\bf r}) \to  {\vec n}({\bf r},t)= (\sin[\pi^a({\bf r})+\alpha\left(t\right)],\cos[\pi^a({\bf r})+\alpha\left(t\right)],0) \ .
\end{equation*}
$\alpha$ depends on time, but it is is space-independent, because it describes a global rotation. The Lagrangian of the system consists of the usual kinetic term, proportional to $({\partial_t{\vec n}})^2$, and the potential energy $E\left({\vec n}\right)$ 
\begin{align} \label{eq:l}
L &=\int d^2 r \left\{ \frac{\chi_\perp}{2} ({\partial_t{\vec n}})^2-E({\vec n}) \right\}\nonumber \\
&\to \int d^2 r \left\{ \frac{\chi_\perp}{2} {\dot{\alpha}}^2-\frac{\rho_s}{2}\left(M_a^2n_a^2+M_c^2n_c^2\right)\right\} \ .
\end{align}
Without anisotropies, i.e. $M_\alpha=0$, the system is rotationally invariant and the potential energy therefore independent of the uniform rotation $\alpha$ (Goldstone theorem). Only the mass terms break the $O(3)$ symmetry and we thus restrict the Lagrangian~(\ref{eq:l}) to these terms. The anisotropy in $c$-direction is irrelevant, because ${\vec n}$ has no c-component. Expanding the $M_a$-term in~(\ref{eq:l}) in powers of $\alpha$ (omitting zeroth order terms), one finds
\begin{equation*}
-\frac{\rho_s}{2}M_a^2\int n_a^2d^2r =-\alpha^2\frac{\rho_s}{2}M_a^2\int\cos[2\pi^a({\bf r})]d^2r \ .
\end{equation*}
The contribution linear in $\alpha$ disappears, because $\int\sin[2\pi^a({\bf r})]d^2r=0$. The Euler-Lagrange equation of motion reads
\begin{align} \label{eq:alphadot}
\Omega \chi_\perp \ddot{\alpha}\left(t\right) &=-\alpha\ \rho_s M_a^2 \int \cos[2\pi^a({\bf r})] d^2 r\nonumber\\ &\approx -\alpha \ \rho_s M_a^2 \Omega(1-2\Pi) \ ,
\end{align}
where $\Omega$ is the area of the sample, and
\begin{multline}  \label{eq:pi2}
\Pi= \frac{1}{\Omega} \int d^2r \ \left[ \pi^a \left({\bf r}\right) \right]^2 
= \frac{1}{\Omega} \int \frac{d^2q}{\left(2 \pi\right)^2}  \left( \pi^a_{\bf q}\right)^2  \\
=x \int \frac{d^2q}{\left(2 \pi\right)^2}  G^a_{\bf q} D^a_{\bf q}\ .
\end{multline}
In order to obtain this last expression, we expanded the right hand side of~(\ref{eq:alphadot}) in powers of the static field $\pi^a$ and also used Eq.~(\ref{eq:crosssection}). The solution of Eq.~(\ref{eq:alphadot}) gives the frequency corresponding to the gap resonance
\begin{equation} \label{eq:gap}
\Delta_{DM}=c M_a \left(1-\Pi\right) \ .
\end{equation}
In the limit of noninteracting dipoles, $D^a_{\bf q}=1$, $\Pi$ can be calculated analytically. Using Eqs.~ (\ref{eq:pi2}), (\ref{eq:g}), and (\ref{eq:rho}) one finds  $\Pi=x\frac{{\cal M}^2}{4\pi}\left(\ln(2\kappa/M_a)-5/4\right)\approx 13x$. However, the interaction of dipoles is very important. Using the correlator $D^a_{\bf q}$ found in Sec.~\ref{sec:correlations} in the numerical integration of~(\ref{eq:pi2}), we find that $\Pi\approx 25.5x$. At $x=0.01$, this gives a 26\% reduction of the gap, in very good agreement with the experimental value of 28\%. Note that $\Pi$ determines the average angle of spin deviation from the $b$-direction, $\overline{\varphi^2}=\Pi$ (see Fig.~\ref{fig:vectorfield2}). Thus at $x=0.01$, the root mean square value of the angle is $\varphi_\text{rms}\approx 0.5\ rad$.

The reduction of the DM spin-wave gap upon doping has recently been considered in Ref.~\onlinecite{juricic05}. Their approach~\cite{juricic05} is based on the introduction of an effective phenomenological Lagrangian for the ``dipolar field''. Our detailed numerical simulations show that the dynamics of the
dipoles are mainly diffusive, so it is {\it impossible} to introduce such an effective Lagrangian. In addition, the authors~\cite{juricic05} suggest a helical spin structure for $0.02 < x < 0.055$. Our results do not support this scenario. At least in the N\'eel phase, the spins remain confined to the $ab$-plane (apart from a small DM canting) and we expect the same for the spin-glass phase.

\section{Uniform magnetic susceptibilities\label{sec:susceptibilities}}
The magnetic susceptibility of undoped LCO, including its temperature dependence is  well understood~\cite{thio88} (see also a recent work by Silva Neto {\it et al.}~\cite{silvaneto05}). There are four mechanisms that contribute to the magnetic susceptibility of LCO: 1)~isotropic atomic core diamagnetism, 2)~anisotropic van Vleck paramagnetism, 3)~anisotropic quantum Heisenberg model paramagnetism, and 4)~anisotropic paramagnetism related to the relativistic DM interaction. In the present work, we consider the doping dependence of the uniform susceptibilities at zero temperature. Due to the anisotropies, there are three different susceptibilities, $\chi_a$, $\chi_b$, and $\chi_c$, corresponding to directions of the magnetic field along the $a$-, $b$-, and $c$-axis, respectively.

The magnetic susceptibilities of lightly doped LSCO have been measured by Lavrov {\it et al.}~\cite{lavrov01}. Fig.~1 of Ref.~\onlinecite{lavrov01} clearly shows that their doping dependence is strongly anisotropic:  $\chi_a$ and $\chi_c$ change only slightly with doping~\cite{comment1}, whereas $\chi_b$ varies from about  $1.7 \cdot 10^{-7} \, emu/g$ at $x=0$ to $4 \cdot 10^{-7} \,Êemu/g$ at $x=0.01$. In comparison, the perpendicular susceptibility of the Heisenberg model is equal to 
\begin{equation*}
\chi_{\perp}\approx 0.5/(8J)\to \frac{(g_s\mu_B)^2}{16J} \to 1.6 \cdot 10^{-7} \,Êemu/g \ .
\end{equation*}
Here we have restored the gyromagnetic ratio, $g_s\approx 2$, and the Bohr magneton $\mu_B$ 
(throughout this paper, we use units where $g_s \mu_B=1$) and substituted the real density of the compound. The variation of the longitudinal susceptibility $\chi_b$ (it is longitudinal because the field is directed along the N\'eel direction) at a tiny 1\% doping is comparable or even slightly larger than $\chi_{\perp}$. Similar to the neutron scattering considered in the previous Secs., this is an indication of a noncollinear spin structure.

In the absence of a magnetic field, the Hamiltonian of the system is given by Eqs.~(\ref{eq:en}), (\ref{eq:epsi}), and (\ref{eq:epsin}). The additional terms describing the interaction with the magnetic field read
\begin{equation} \label{eq:eb}
E_B = E_{Bn}+E_{B\psi}^{(1)}+E_{B\psi}^{(2)} \ ,
\end{equation}
with
\begin{align} \label{eq:ebnpsi}
E_{Bn} &= -\frac{\chi_{\perp}}{2} \int d^2r \left[ \left( {\vec n}\times{\vec B} \right)^2  -2{\vec B} \cdot \left({\vec D}_+\times {\vec n}\right) \right] \ , \nonumber\\
E_{B\psi}^{(1)} &=-\frac{1}{2}\int d^2r  \left(\psi^{\dag} {\vec \sigma}\cdot{\vec n}\psi\right) ({\vec B}\cdot{\vec n}) \ , \nonumber \\
E_{B\psi}^{(2)} &=\frac{g}{2\sqrt{2}} \int d^2r  \psi^\dag {\vec \sigma} \left\{ {\bf e} \cdot {\bf p}, 
{\vec B}-{\vec n} \left( {\vec B} \cdot {\vec n} \right) \right\} \psi \ ,
\end{align}
where $\{,\}$ stands for the anticommutator, ${\bf p}=-i {\bf \nabla}$ and we set $g_s\mu_B=1$.  Note that the field ${\vec n}$ and the spinor $\psi$ are functions of ${\bf r}$. The first term in $E_{Bn}$ is the usual magnetic interaction in the NLSM~\cite{chakkravarty88,fisher89} 
and  the second term is due to the DM induced weak ferromagnetism~\cite{thio88,silvaneto05}.

$E_{B\psi}^{(1)}$ describes the interaction between the hole and the component of the magnetic field parallel to the local direction of the ${\vec n}$-field. The physical origin of this term is very simple. In the Heisenberg model (undoped case), the projection of the total spin along ${\vec n}$ is zero, because electrons with spin ``up'' are  compensated by electrons with spin ``down''. For nonzero doping, ``up'' and ``down''-spins no longer compensate and therefore lead to a nonzero spin-projection along ${\vec n}$. This spin interacts in the usual (Zeeman) way with the magnetic field. For example, $\Psi^{\dag}{\vec \sigma}\cdot{\vec n}\Psi=+1$ implies that the hole is created on the $\uparrow$-sublattice and hence $\Delta S_n=+1$.
 
The origin of $E_{B\psi}^{(2)}$ is more delicate. This term describes the interaction between the  hole and the component of the magnetic field orthogonal to the local direction of ${\vec n}$. The physical origin of this interaction is the following: the locally transverse magnetic field tilts the spins in the antiferromagnetic background, which allows the holes to hop more easily and therefore leads to a gain in kinetic energy. This is why the $E_{B\psi}^{(2)}$-term contains the momentum, ${\bf p}=-i\nabla$, and is proportional to the hopping matrix element $g=Zt$. To derive the coefficient in $E_{B\psi}^{(2)}$, one has to calculate the gain in the kinetic energy and then expand it for a hole localized in a particular pocket, ${\bf q}=(\pm \pi/2,\pm\pi/2)+{\bf p}$. This why the unit vector ${\bf e}$, orthogonal to the corresponding face of the MBZ appears in this expression. However, because in the problems we consider, momenta are always small, ${\bf p} \ll 1$, we can safely neglect the $E_{B\psi}^{(2)}$-term. We nevertheless present the term in Eqs.~(\ref{eq:eb}) and (\ref{eq:ebnpsi}) for the sake of completeness and for future studies.

To the best of our knowledge, the interaction between the hole and the magnetic field has never been written in the form~(\ref{eq:ebnpsi}) before. It is therefore useful to convince oneself of its validity by recalculating a well known result. A nice example is the calculation of the perpendicular susceptibility of a single immobile hole in a quantum antiferromagnet~\cite{sachdev03,sushkov03}. Since the hole is located on a particular site of the lattice, 
$\psi^{\dag}{\vec \sigma}\cdot{\vec n}\psi=\pm\delta\left({\bf r}\right)$, and because $g=0$
(no hopping), the total energy reads
\begin{equation*}
\frac{\rho_s}{2} \int d^2 r \left\{ \left[{\bf \nabla}{\vec \theta}\left({\bf r}\right) \right]^2 +M^2 \left[{\vec \theta}^2\left({\bf r}\right)\right]^2
\mp \frac{1}{\rho_s}{\vec B}\cdot{\vec \theta}\left({\bf r}\right) \delta({\bf r}) \right\} \ .
\end{equation*}
Here the terms in square brackets are due to Eq.~(\ref{eq:etot}), and the last term is due to $E_{B\psi}^{(1)}$ in~(\ref{eq:ebnpsi}). Variation of the energy with respect to ${\vec \theta}$ yields
\begin{equation*}
\left(-\triangle+M^2\right) {\vec \theta}\left({\bf r}\right)=  \pm \frac{{\vec B}}{2\rho_s} \delta\left({\bf r}\right) \ ,
\end{equation*}
which in momentum representation has the solution
\begin{equation*}
{\vec \theta}_{\bf q} = \pm \frac{{\vec B}}{2\rho_s} \frac{1}{q^2+M^2} \ .
\end{equation*}
Substituting this solution in the total energy and using the ultraviolet  regularization $r \ge 1$  lattice spacing, i.e. $q \le 1$, then immediately yields the ``impurity'' susceptibility
\begin{equation*}
\chi_\text{imp} = \frac{1}{8 \pi \rho_s} \ln\left(\frac{1}{M}\right) \ ,
\end{equation*}
which agrees with previous results obtained by different methods~\cite{sachdev03,sushkov03}.

Let us now turn to the realistic situation of LSCO, where mobile holes are trapped by Sr ions (``impurities''). To calculate the susceptibility, it is sufficient to consider a single ``impurity'' and then multiply  the result by the concentration $x$, because the interaction of ``impurities'' considered in Secs.~\ref{sec:model} and \ref{sec:correlations} gives corrections of second and higher order in doping, which we cannot calculate without uncontrolled approximations, see also comment below Eq.~(\ref{eq:chib}). In what follows, we set $M_a=0$ everywhere except in the logarithmically divergent integrals, where we use it as an infrared cutoff. Accordingly, the term involving ${\vec D}_+$ in Eq.~(\ref{eq:ebnpsi}) is also ignored.

\subsection{Longitudinal magnetic field, $\chi_b$}
In the case of a longitudinal magnetic field ${\vec B} = B {\vec e}_{b}$, the ``impurity'' energy is given by Eqs.~(\ref{eq:etot}), (\ref{eq:eb}), and (\ref{eq:ebnpsi})
\begin{multline}  \label{eq:etotlong}
E = \frac{1}{2} \int d^2 r \bigg\{ \rho_s \left[ {\bf \nabla} {\vec \theta}\left({\bf r}\right)\right]^2 \bigg. \\
\bigg. + 2\sqrt{2}g\rho({\bf r})\left[{\vec e}_b\times{\vec d}\,\right]\left({\bf e}_b\cdot{\bf \nabla}\right){\vec \theta}\left({\bf r}\right) \bigg. \\ 
-\bigg. \chi_\perp B^2 \left[{\vec \theta}\left({\bf r}\right)\right]^2  - B\left[ d^\parallel + {\vec d}^\perp \cdot {\vec \theta}\left({\bf r}\right) \right] \rho \left({\bf r}\right)  \bigg\} \ ,
\end{multline}
where we represent the vector ${\vec d} = \Psi^\dag {\vec \sigma} \Psi$ as  ${\vec d} = d^\parallel {\vec e}_{b} + {\vec d}^\perp$. Variation of the energy with respect to ${\vec \theta}$ yields
\begin{equation*}
-\triangle {\vec \theta} \left({\bf r}\right) -\left[ {\cal M} \left({\vec e}_b \times {\vec d}^\perp\right)  \left({\bf e}_b\cdot{\bf \nabla}\right)+\frac{B {\vec d}^\perp}{2 \rho_s} \right] \rho\left({\bf r}\right) = 0 \ . 
\end{equation*} 
In momentum representation, the solution of this equation reads
\begin{equation} \label{eq:thetaBlong}
{\vec \theta}_{\bf q} = \left[i {\cal M}\left({\vec e}_b \times {\vec d}^\perp\right)   {\bf e}_b \cdot {\bf q} + \frac{B {\vec d}^\perp}{2\rho_s} \right] \frac{\rho_{\bf q} }{q^2} \ ,
\end{equation}
where the first term, which is independent of the magnetic field $B$, coincides with~(\ref{eq:theta}). Note that analogous to neglecting the mass terms, we also omit the $\chi_{\perp} B^2 \theta^2$ contribution in the solution~(\ref{eq:thetaBlong}), because it only leads to a small modification of the infrared cut-off. Substitution of this solution in the total energy~(\ref{eq:etotlong}) yields
\begin{multline} \label{eq:Ei}
E = - \left(d^\perp\right)^2  \left[ \frac{B^2}{4 \pi \rho_s} \left( \frac{1}{4}+\frac{\chi_\perp g^2}{\rho_s} \right)  \ln \left(\frac{0.6 \kappa}{M_a}\right)  \right. \\
\left. +\frac{g^2}{4\pi\rho_s} \kappa^2 \right] - d^\parallel \frac{B}{2} \ ,
\end{multline} 
with the kinematical constraint $(d^\parallel)^2+(d^{\perp})^2=1$. The energy is minimized for a non-zero $d^\parallel$, i.e., a longitudinal magnetic field leads to a squeezing of ${\vec m}={\vec e}_b \times {\vec d}$.  The minimum is obtained for 
\begin{equation*}
d_\parallel \approx \frac{\pi \rho_s}{g^2 \kappa^2} B \ .
\end{equation*}
Substitution of this expression in~(\ref{eq:Ei}) gives the ``impurity'' energy and allows to find the ``impurity'' susceptibility. After multiplication by the hole concentration $x$, we find the following variation of the bulk susceptibility
\begin{align} \label{eq:chib} 
\delta\chi_b &=\left[ \left( \frac{c^2}{8\pi \rho_s^2} + \frac{g^2}{2\pi\rho_s^2} \right) 
\ln\left(\frac{0.6\kappa}{M}\right)+ \frac{\pi c^2}{2 g^2\kappa^2}
\right] x \chi_{\perp} \nonumber \\
&\sim 100 x \chi_{\perp} \ .
\end{align}
The last term in the above expression is the most important one. Depending on the value of $\kappa$, it gives 80\% - 90\% of the total result. From data on hopping conductivity~\cite{chen95,panagopoulos05}, we know that $0.2 \lesssim \kappa \lesssim 0.3$. Evaluation of Eq.~(\ref{eq:chib}) for $\kappa=0.3$ gives $\delta\chi_b \approx 75x \chi_{\perp}$ and with $\kappa=0.2$, we find $\delta\chi_b \approx 130x \chi_{\perp}$. Similar to the uncertainty in $\kappa$, there is also some variability in the coupling constant $g$. According to Refs.~\onlinecite{igarashi92,sushkov04}, we take $g\approx 1$, but we believe that $g\approx 1\pm 0.2$ is quite possible. Despite these uncertainties, our result unambiguously shows that there is a huge doping dependent variation of the longitudinal susceptibility due to the noncollinear spin structure of the spiral. Our estimation of $\delta\chi_b$ is in excellent agreement with experimental data~\cite{lavrov01}. We would like to emphasize that Eq.~(\ref{eq:chib}) has been derived in the fully controlled linear in $x$ approximation. Unfortunately, higher order doping terms cannot be calculated without uncontrolled approximations. Eq.~(\ref{eq:chib}) is therefore justified when $x$ is well below the transition to the spin-glass phase at $x=0.02$. Since our results on neutron and Raman scattering are also derived in the linear in $x$ approximation, and we know from comparison with experiments that they are valid up to $x\approx 0.015$, we expect Eqs.~(\ref{eq:chib}) and (\ref{eq:chiac}) to be valid in this region as well. In contrast to this huge modification of the longitudinal susceptibility upon doping, practically no variation of the transverse susceptibilities has been observed experimentally~\cite{lavrov01}.

\subsection{Transverse magnetic field, $\chi_a$ and $\chi_c$}
For a transverse magnetic field ${\vec B} \perp {\vec e}_{b}$, the ``impurity'' energy  given by Eqs.~(\ref{eq:etot}), (\ref{eq:eb}), and (\ref{eq:ebnpsi}) reads
\begin{multline} \label{eq:etottrans}
E = \frac{1}{2} \int d^2 r \bigg\{ \rho_s \left[{\bf \nabla}{\vec \theta} \left({\bf r}\right)\right]^2 \bigg. \\ \bigg. +
2\sqrt{2} g \left({\vec e}_b\times{\vec d}\,\right) \rho\left({\bf r}\right) \left({\bf e}_b\cdot {\bf \nabla}\right){\vec \theta} \left({\bf r}\right) \bigg. \\
+ \bigg. \chi_\perp \left[ {\vec B} \cdot {\vec \theta} \left({\bf r}\right)\right]^2 -\left[{\vec B} \cdot {\vec \theta}\left({\bf r}\right)\right] \left[{\vec d}^\perp \cdot {\vec \theta}\left({\bf r}\right)\right]  \rho\left({\bf r}\right) \bigg\} \ .
\end{multline}
In this case, $d^\parallel=0$ and ${\vec d} ={\vec d}^\perp$, which maximizes the ``impurity'' dipole moment and hence minimizes the energy. Let us write the ${\vec \theta}$-field as
\begin{equation} \label{eq:theta2}
{\vec \theta}\left({\bf r}\right)={\vec \theta}^{\,(0)}\left({\bf r}\right)+{\vec \theta}^{\,(1)}\left({\bf r}\right) \ ,
\end{equation}
where ${\vec \theta}^{\,(0)}$ is the solution for ${\vec B}=0$ given by Eq.~(\ref{eq:theta}) and  ${\vec \theta}^{\,(1)}$ is the perturbation induced by the magnetic field. Performing the variation of the energy (\ref{eq:etottrans}) with respect to ${\vec \theta}^{\,(1)}$ yields the equation
\begin{multline*}
-\triangle{\vec \theta}^{\,(1)}\left({\bf r}\right) - \\
\frac{1}{2\rho_s} \left\{ {\vec B} \left[ {\vec d}^\perp \cdot {\vec \theta}\left({\bf r}\right)\right]+ 
{\vec d}^\perp \left[ {\vec B} \cdot {\vec \theta}\left({\bf r}\right) \right]  \right\} \rho\left({\bf r}\right) = 0 \ .
\end{multline*}
Using the explicit form of ${\vec \theta}^{\,(0)}$ given by Eq.~(\ref{eq:theta}) (we also set $M_\alpha=0$) we then find the magnetic field induced part of the spiral
\begin{multline*} 
{\vec \theta}^{\,(1)}_{\bf q} = -i \frac{\kappa^2 {\cal M}}{2\pi\rho_s} {\vec d} \left( {\vec B} \cdot {\vec m}\right) \frac{Ê{\bf e}_b \cdot {\bf q}}{q^4} \\ \left[ \frac{2\kappa}{\sqrt{4 \kappa^2+q^2}} +
\frac{2 \kappa q^2}{\left(16 \kappa^2+q^2\right)^{3/2}}-\frac{4\kappa}{\sqrt{16\kappa^2+q^2}}\right] \ .
\end{multline*}
Substitution of this solution together with~(\ref{eq:theta}) in Eq.~(\ref{eq:theta2}) and then  in Eq.~(\ref{eq:etottrans}) gives the total energy and hence the ``impurity'' susceptibility. After multiplication by the hole concentration $x$, we find the variation of the bulk susceptibilities
\begin{align} \label{eq:chiac}
\delta \chi_a &= \left[ -\frac{\chi_\perp {\cal M}^2}{4 \pi} \ln\left( \frac{0.6\kappa}{M_a} \right)
+ 0.0043\frac{\kappa^2 {\cal M}^2}{4 \pi^3 \rho_s} \right] x \nonumber \\ &
\sim -10x \chi_{\perp} \ , \nonumber \\
\delta \chi_c &= 0.0043\frac{\kappa^2 {\cal M}^2}{4 \pi^3 \rho_s} x \sim 0.02 x \chi_\perp \ ,
\end{align}
where we have used the fact that ${\vec \theta} \propto {\vec e}_{a}$. The term $\left( {\vec B} \cdot {\vec \theta}\,\right)^2$ in~(\ref{eq:etottrans}) is thus only present for a magnetic field directed along the $a$-axis. Such a small variation of the susceptibilities upon doping is quite consistent with experiments~\cite{lavrov01}.

\section{Conclusion\label{sec:conclusion}}
In the present work based on the spiral solution of the extended $t$-$J$ model, we explained the following properties of underdoped La$_{2-x}$Sr$_x$CuO$_4$.\\
1) The pinning of the incommensurate magnetic structure to the orthorhombic $b$-direction observed in neutron scattering in the insulating phase, $x< 0.055$. The pinning is due to the anisotropy of the diagonal hopping matrix element $t'$, see Sec.~\ref{sec:pinning}.\\
2) The positions and shapes of the incommensurate elastic neutron scattering peaks in the N\'eel phase, $x<0.02$. Experimental data are presented in Fig.~\ref{fig:neutronscattering}, together with our theoretical curves, containing no fitting parameters at all. The agreement between theory and experiments is quite remarkable.\\
3) The doping dependence of the Dzyaloshinski-Moriya induced spin-wave gap in the N\'eel phase, see Eq.~(\ref{eq:gap}). According to our calculation, at 1\% doping, the gap is reduced by 26\% compared to its value in the undoped compound. This is in very good agreement with the experimentally observed reduction of 28\%.\\
4) The doping dependence of the uniform magnetic susceptibilities at zero temperature, see Eqs.~(\ref{eq:chib}) and (\ref{eq:chiac}). This explains the tremendous variation of the longitudinal magnetic susceptibility $\chi_b$ and the very weak change in the transverse susceptibilities $\chi_a$ and $\chi_c$.

\acknowledgments{We are grateful to Y. Ando, V.~N. Kotov, and A.~N. Lavrov for stimulating discussions and would like to thank M.~Matsuda for providing us with the experimental data. G.~M. and A.~I.~M. gratefully acknowledge the School of Physics at the University of New South Wales for warm hospitality and financial support during their visit. G.~M. is in part supported by the French Minist\`ere de la Recherche et des Nouvelles Technologies with an ACI grant.}


\begin{thebibliography}{99}
\bibitem{comment1} We disregard a very weak Curie-like behaviour observed for both $x=0$ and $x=0.01$.
\end{thebibliography}
\end{document}